\documentclass[conference]{IEEEtran}
\IEEEoverridecommandlockouts
\usepackage{amsmath}
\usepackage{cite}
\usepackage{graphicx}
\usepackage{float}
\usepackage{times}
\usepackage{latexsym}
\usepackage{bm}
\usepackage{amssymb}
\usepackage[center]{caption2}
\usepackage{stfloats}
\usepackage{array}
\usepackage{fancyhdr}
\usepackage{cite,graphicx,amssymb}
\usepackage{citesort}
\usepackage{psfrag}
\usepackage{multirow}

\usepackage{color}

\ifCLASSINFOpdf

\else

\fi

    \def\Complex{{\rm\rule[.23ex]{.03em}{1.1ex}\kern-.3em{C}}}

    \newcommand{\be}{\begin{equation}} \newcommand{\ee}{\end{equation}}
    \newcommand{\bea}{\begin{eqnarray}} \newcommand{\eea}{\end{eqnarray}}
    \newcommand{\benum}{\begin{enumerate}} \newcommand{\eenum}{\end{enumerate}}

    \newcommand{\qd}{{\bf d}}

    \newcommand{\qn}{{\bf n}}

    \newcommand{\qx}{{\bf x}}
    \newcommand{\qy}{{\bf y}}
    \newcommand{\qz}{{\bf z}}

    \newcommand{\qA}{{\bf A}}
    \newcommand{\qB}{{\bf B}}

    \newcommand{\qH}{{\bf H}}
    \newcommand{\qI}{{\bf I}}

    \newcommand{\qU}{{\bf U}}
    \newcommand{\qV}{{\bf V}}
    \newcommand{\qW}{{\bf W}}

    \newcommand{\qXi}{{\boldsymbol \Xi}}

    \newcommand{\qOmega}{{\boldsymbol \Omega}}

    \newcommand{\bbC}{{\mathbb C}}

    \newcommand{\tr}{{\sf tr}}

\newtheorem{alg}{Algorithm}

\newtheorem{example}{Example}

\makeatother

\begin{document}
\title{Low-Complexity MIMO Precoding with Discrete Signals and Statistical CSI}

\author{\IEEEauthorblockN{Yongpeng Wu, Chao-Kai Wen, Derrick Wing Kwan Ng, Robert Schober, and Angel Lozano}

\thanks{The work of Y. Wu and R. Schober was supported by the Alexander von
Humboldt Foundation. The work of C.-K. Wen was supported in part by the the Ministry of Science and Technology,
Taiwan, under Grant MOST103-2221-E-110-029-MY3.}

\thanks{Y. Wu and R. Schober are with Institute for Digital Communications, Universit\"{u}t Erlangen-N$\ddot{u}$rnberg,
Cauerstrasse 7, D-91058 Erlangen, Germany (Email: yongpeng.wu@fau.de;  schober@fau.de).}

\thanks{D. W. K. Ng is with the School of Electrical Engineering and
Telecommunications, University of New South Wales, Sydney, N.S.W.,
Australia (E-mail: w.k.ng@unsw.edu.au).}

\thanks{C. K. Wen is with the Institute of Communications Engineering, National Sun Yat-sen University, Kaohsiung 804,
Taiwan (Email:  chaokaiwen@gmail.com).}

\thanks{A. Lozano is with Dept. of Information and Communication Technologies, Universitat Pompeu Fabra, C/Roc Boronat 138,
08018, Barcelona, Spain (Email: angel.lozano@upf.edu).}

}

\maketitle

\begin{abstract}
In this paper, we investigate the design of multiple-input multiple-output single-user precoders for finite-alphabet signals
under the premise of statistical channel-state information at the transmitter.
Based on an asymptotic expression for the mutual information of channels exhibiting antenna correlations, we propose a low-complexity iterative algorithm that radically reduces the computational load of existing approaches by orders of magnitude with only minimal losses in performance.
The complexity savings increase with the number of transmit antennas and with the cardinality of the signal alphabet, making it possible to support values thereof that were unwieldy in existing solutions.
\end{abstract}


\section{Introduction}

Although Gaussian signals are capacity-achieving in a multiple-input multiple-output (MIMO) channel under perfect channel-state information (CSI) at the receiver, signals conforming to discrete constellations are transmitted in practice, and the design of precoders optimized for such signal formats is a topic that has gathered momentum in recent years \cite{Lozano2006TIT,Lozano2008TCOM,Perez-Cruz2010TIT,Xiao2011TSP,Mohammed2011TIT,Wu2012TVT,Wu2012TWC,Wu2013TCOM,Ketseoglou2015TWC,Ketseoglou2016}.

The works in \cite{Perez-Cruz2010TIT,Xiao2011TSP,Mohammed2011TIT,Wu2012TVT,Wu2012TWC,Wu2013TCOM,Ketseoglou2015TWC,Ketseoglou2016} consider the problem under the assumption of perfect CSI at the transmitter, which is a reasonable premise in reciprocal or slow fading channels. Often though, perfect CSI at the transmitter is an impossibility and only statistical CSI is available therein; these are the conditions on which we concentrate here. For Gaussian signals, the design of MIMO precoders with statistical CSI has been addressed in \cite{tulino2005TIT,tulino2006capacity,Gao,wen2011sum,JWang2012TSP,Zhang2013JSAC,Wu2014TSP,Wu2015}. For discrete signals, an iterative precoding algorithm was proposed in \cite{zeng2012linear} and shown to achieve a high ergodic spectral efficiency in simulations. However, the complexity of this complete-search algorithm is exponential in the number of transmit antennas and, even with modest numbers thereof (say, eight), it becomes unwieldy.

The alternative algorithm proposed in this paper drastically reduces the search space, and with it the complexity, but in such a way that the loss in performance---established based on the 3GPP spatial channel model (SCM) \cite{Salo2005} ---is minimal.

The remainder of this paper is organized as follows. Section II describes the system model.
In Section III, we review the complete-search algorithm and propose an idea to
reduce its computational complexity.  In Section IV, we  propose
a low complexity precoder design. Numerical results are provided in Section V, and our main results are summarized in Section VI.

The following notations are adopted throughout the paper:
 ${\rm{diag}}\left\{\bf{A}\right\}$  denotes a diagonal matrix containing the diagonal of matrix $\mathbf{A}$, ${\rm vec} \left(\mathbf{A}\right) $ is a column vector containing the stacked columns of matrix $\mathbf{A}$,
$[\mathbf{A}]_{mn}$ denotes the $(m,n)$th entry of matrix $\mathbf{A}$,  $[\mathbf{a}]_{m}$ denotes the $m$th entry
of vector $\mathbf{a}$, ${\mathbf{I}}_M$ denotes an $M \times M$ identity matrix, $\rm{tr}(\cdot)$ denotes the trace operation,
$\rm{det}(\cdot)$ denotes the matrix determinant, and $E_V\left[\cdot \right]$ represents the expectation with respect to random variable $V$, which can be a scalar, vector, or matrix.

\section{Signal Model}
\label{sec:model}

Consider a single-user MIMO channel where transmitter and receiver
are equipped with $N_{\mathrm{t}}$ and $N_{\mathrm{r}}$ antennas, respectively. The received signal
$\qy\in{\mathbb C}^{N_{\mathrm{r}}}$ can be
written as
\begin{equation}\label{eq:x}
 \qy = \qH \qx + \qn
\end{equation}
where $\qH \in{\mathbb C}^{N_{\mathrm{r}} \times N_{\mathrm{t}}}$ is a random channel matrix whose $(i,j)$th entry denotes the complex fading coefficient between the $j$th transmit and the $i$th receive antenna, $\qx \in{\mathbb C}^{N_{\mathrm{t}}}$  denotes
zero-mean transmitted vector with covariance $\boldsymbol{\Sigma}_{\qx}$, and
$\qn \in{\mathbb C}^{N_{\mathrm{r}}}$ is a zero-mean
complex Gaussian noise vector with covariance $\qI_{N_{\mathrm{r}}}$.
The transmit vector $\qx$ satisfies the power constraint
\begin{equation}\label{x_constraint_2}
{{\rm{tr}}\left(\boldsymbol{\Sigma}_{\qx} \right)} \leq P.
\end{equation}%
Based on the statistical CSI, and subject to the power constraint, the transmitter needs to optimize $\boldsymbol{\Sigma}_{\qx}$
to maximize the ergodic spectral efficiency.

With $\qH$ known at the receiver, the ergodic mutual information between $\qx$ and $\qy$  is given by \cite{Cover}
\begin{equation}\label{eq:Mutual_Info}
I(\qx;\qy)=E \left[E \left[\left.\log\frac{p (\qy|\qx,\qH)} {p(\qy|\qH)}\right|\qH\right]\right].
\end{equation}
In (\ref{eq:Mutual_Info}), $p(\qy|\qx,\qH)$ denotes
the probability density function (p.d.f.) of
$\qy$ conditioned on $(\qx,\qH)$, and ${p(\qy|\qH)}$
denotes the p.d.f of $\qy$ conditioned on $\qH$.


\section{Complete-Search Precoder Design} \label{sec:transmit_design}
In this section, we review the complete-search approach for optimization of
$\boldsymbol{\Sigma}_{\qx}$ and introduce an idea to reduce its computational load
in the case where instantaneous CSI is available
at the transmitter. Then, we will extend this idea to the case where only statistical CSI is available at the transmitter in next section.

Let $\qx = {{{\bf{B}}} \,{{\bf{d}}}}$, where ${{\bf{d}}} \in \mathbb{C}^{N_{\mathrm{t}} \times 1}$
is the signal vector drawn from an equiprobable constellation of size $M^{N_{\mathrm{t}}}$ whereas $\mathbf{B} \in \mathbb{C}^{N_{\mathrm{t}} \times N_{\mathrm{t}}}$ is the precoder.
Let ${\mathbf{d}}_{m}$ denote the $m$th element in the constellation.
Consider the singular value decomposition (SVD)
$\mathbf{B} = \mathbf{U}_{{\mathrm{\bf{B}}}} \boldsymbol{\Lambda}_{{\mathrm{\bf{B}}}} \mathbf{V}_{{\mathrm{\bf{B}}}}$
where $\boldsymbol{\Lambda}_{{\mathrm{\bf{B}}}} \in \mathbb{C}^{N_{\mathrm{t}} \times N_{\mathrm{t}}}$
is diagonal while $\mathbf{U}_{{\mathrm{\bf{B}}}} \in \mathbb{C}^{N_{\mathrm{t}} \times N_{\mathrm{t}}}$ and $\mathbf{V}_{{\mathrm{\bf{B}}}} \in \mathbb{C}^{N_{\mathrm{t}} \times N_{\mathrm{t}}}$ are unitary.

When Gaussian-signal precoding solutions are applied to discrete constellations, the performance suffers because,
in the face of major power variations between MIMO subchannels, these solutions insist on beamforming over an extensive range of signal-to-noise ratios (SNRs), well beyond the point where beamforming is appropriate for a discrete constellation.
With beamforming, signalling is only possible over the dominant subchannel, which causes a performance loss with discrete signals (cf. \cite{Xiao2011TSP,zeng2012linear}).
By properly designing $\mathbf{U}_{{\mathrm{\bf{B}}}}$, $\boldsymbol{\Lambda}_{{\mathrm{\bf{B}}}}$, and $\mathbf{V}_{{\mathrm{\bf B}}}$,
the complete-search precoder design minimizes this loss \cite{Xiao2011TSP,zeng2012linear}.
Thereby, the matrix $\mathbf{V}_{{\mathrm{\bf{B}}}}$ mixes the $N_{\mathrm{t}}$ original signals into $N_{\mathrm{t}}$ beams, then $\boldsymbol{\Lambda}_{{\mathrm{\bf{B}}}}$ allocates power to those beams, and finally $\mathbf{U}_{{\mathrm {\bf{B}}}}$ aligns them spatially as they are launched onto the channel.
With a proper choice of $\mathbf{V}_{{\mathrm{\bf{B}}}}$, all the $N_{\mathrm{t}}$ signals can be
effectively transmitted even if only a single beam is active.

The following  example illustrates the role of $\mathbf{U}_{{\mathrm{\bf{B}}}}$, $\boldsymbol{\Lambda}_{{\mathrm{\bf{B}}}}$, and $\mathbf{V}_{{\mathrm{\bf{B}}}}$.

\begin{example} Consider a $4 \times 4$ deterministic channel ${{\bf{A}}}$ with SVD
${{\bf{A}}} = \mathbf{U}_{{\mathrm{\bf{A}}}} \boldsymbol{\Lambda}_{{\mathrm{\bf{A}}}} \mathbf{V}_{{\mathrm{\bf{A}}}}$, which is perfectly known
at the transmitter. The received signal is given by
\begin{equation}
\label{eq:y_4_exp}
{\bf{y}} = {\bf{A}} \, {{\bf{U}}_{{{\mathrm{\bf{B}}}}}}{{\bf{\Lambda }}_{{{\mathrm{\bf{B}}}}}}{{\bf{V}}_{{{\mathrm{\bf{B}}}}}} \, {\bf{d}} + {\bf{n}}
\end{equation}
where ${\bf{d}} = [d_1,d_2,d_3,d_4]^T $.
From \cite[Prop. 2]{Xiao2011TSP}, the optimal design satisfies
$\mathbf{U}_{{\mathrm{\bf{B}}}} = \mathbf{V}_{{\mathrm{\bf{A}}}}^H$. Then, based on \cite[Eq. (8)]{Xiao2011TSP},
(\ref{eq:y_4_exp}) can be rewritten as
\begin{equation}\label{eq:y_4_exp_2}
\begin{array}{l}
\overline {\bf{y}} =
\left[ {\begin{array}{*{20}{c}}
{{a_1\lambda_1}}&{}&{}\\
{}& \ddots &{}\\
{}&{}&{{a_4 \lambda_4}}
\end{array}} \right] \! \! \left[ {\begin{array}{*{20}{c}}
{{V_{11}}}& \hdots &{{V_{14}}}\\
\vdots & \ddots & \vdots \\
{{V_{41}}}&\hdots&{{V_{44}}}
\end{array}} \right]   \mathbf{d} + {{\bf{n}}}
\end{array}
\end{equation}
where $\overline {\bf{y}} = \mathbf{U}_{{\mathrm{\bf{A}}}}^H {\bf{y}}$ while $a_i$
and $\lambda_i$ are the diagonal entries of $\boldsymbol{\Lambda}_{{\mathrm{\bf{A}}}}$
and $\boldsymbol{\Lambda}_{\mathrm{\bf{B}}}$, respectively, and $V_{ij} =\left[\mathbf{V} \right]_{ij}$.

Assume two of the subchannel gains, say $a_2$ and $a_4$, are very weak.
Then, with a Gaussian-signal precoder, the powers allocated to the corresponding subchannels will be
very small even at moderate SNRs. Since, with Gaussian signals, ${{\bf{V}}_{\mathrm{\bf{B}}}}$ is immaterial, $d_2$ and $d_4$ then
cannot be transmitted.
With a proper ${{\bf{V}}_{\mathrm{\bf{B}}}}$, in contrast, the received signal equals
\begin{equation}\label{eq:y_4_exp_3}
{\left[ {\overline {\bf{y}} } \right]_i} = {a_i}{\lambda _i}\sum\limits_{j = 1}^4 {{V_{ij}}{d_j}} \qquad i = 1,2,3,4
\end{equation}
and now, even if $a_2 \lambda_2 \approx 0$ and $a_4 \lambda_4 \approx 0$, $d_2$ and $d_4$ can still
be effectively transmitted along other subchannels.
\end{example}

As indicated by (\ref{eq:y_4_exp_3}), an adequate design for discrete constellations in general mixes
all the signals ($d_1$,$d_2$,$d_3$,$d_4$) and transmits the ensuing beams on different subchannels.
As a result, the search space for computing the mutual information with finite alphabet inputs grows exponentially with $N_{\mathrm t}$ \cite{Xiao2011TSP}.

Intuitively though, if there are only two weak subchannels, say $a_2$ and $a_4$ in Example 1, it is not necessary to mix all the signals.
It suffices to mix $d_2$ with $d_1$ and $d_4$ with $d_3$ and transmit the ensuing beams on the stronger subchannels $a_1$ and $a_3$.
This corresponds to
\begin{equation}\label{eq:y_4_exp_4}
{\bf{V}} = \left[ {\begin{array}{*{20}{c}}
{{V_{11}}}&{{V_{12}}}&0&0\\
{{V_{21}}}&{{V_{22}}}&0&0\\
0&0&{{V_{33}}}&{{V_{34}}}\\
0&0&{{V_{43}}}&{{V_{44}}}
\end{array}} \right]
\end{equation}
which, plugged into (\ref{eq:y_4_exp_2}), gives
\begin{align}
{\left[ {\overline {\bf{y}} } \right]_i} & = {a_i}{\lambda _i}\sum\limits_{j = 1}^2 {{V_{ij}}{d_j}} \qquad  i = 1,2 \label{eq:y_4_exp_5} \\
{\left[ {\overline {\bf{y}} } \right]_i} & = {a_i}{\lambda _i}\sum\limits_{j = 3}^4 {{V_{ij}}{d_j}} \qquad  i = 3,4 .  \label{eq:y_4_exp_6}
\end{align}
Observe from (\ref{eq:y_4_exp_5}) and (\ref{eq:y_4_exp_6}) that
$(d_1,d_2)$ and $(d_3,d_4)$ are decoupled.  If $\mathbf{d}$ is
drawn from quadrature phase shift keying (QPSK) distributions, then the search space for computing
the mutual information with finite alphabet inputs in (\ref{eq:y_4_exp_5}) and (\ref{eq:y_4_exp_6})
is of dimension $2 \times 4^{2 \times 2} = 512$ \cite{Xiao2011TSP}. In contrast, for the complete search in (\ref{eq:y_4_exp_3}), it is of dimension $ 4^{2 \times 4} = 65536$.
Since $d_2$  and $d_4$ are transmitted all the same, the structure in (\ref{eq:y_4_exp_4}) may
perform close to the complete-search design, but with a substantially lower computational complexity.
This observation is the basis for the low-complexity precoder design proposed in the next section.

\section{Low-Complexity Precoder Design}
In this section, we extend the idea above to the case where only statistical CSI is available at the transmitter.
First, we introduce the channel model. Then, we provide an asymptotic (large system limit)
expression of the erogdic mutual information in (\ref{eq:Mutual_Info}). Based on this asymptotic expression, we study
precoder structures, based on  a low complexity numerical algorithm is proposed to design the precoder.

\subsection{Channel Model}
Inspired by (\ref{eq:y_4_exp_5}) and (\ref{eq:y_4_exp_6}), we propose a low-complexity
design to maximize the ergodic spectral efficiency in (\ref{eq:Mutual_Info}).
Thereby, we consider the popular Kronecker channel model \cite{shiu2000fading}
\begin{equation}\label{eq:Spatial_Cov}
\qH = \qA_{{\rm R}}^{1/2}  \qW \,  \qA_{{\rm T}}^{1/2}
\end{equation}
where $\qA_{{\rm R}} \in\mathbb{C}^{N_{\mathrm r} \times N_{\mathrm r}} $ and $\qA_{{\rm T}}
\in\mathbb{C}^{N_{\mathrm t} \times N_{\mathrm t}}$ are transmit and receive correlation matrices while
$\qW\in\mathbb{C}^{N_{\mathrm r} \times N_{\mathrm t}}$ is a random matrix whose entries are independent and identically distributed (IID) complex Gaussians. The eigenvalue decompositions of $\qA_{{\rm R}}$
and $\qA_{{\rm T}}$ are
\begin{align}\label{eq:qA_R}
\qA_{{\rm R}} & = \qU_{\rm R}  \boldsymbol{\Lambda}_{{\rm R}} \qU_{\rm R}^H  \\
\qA_{{\rm T}} & = \qU_{\rm T}  \boldsymbol{\Lambda}_{{\rm T}} \qU_{\rm T}^H
\end{align}
where $\qU_{\rm R} \in\mathbb{C}^{N_{\mathrm r} \times N_{\mathrm r}} $ and $\qU_{\rm T} \in\mathbb{C}^{N_{\mathrm t} \times N_{\mathrm t}} $
are  unitary matrices, and $\boldsymbol{\Lambda}_{{\rm R}} \in\mathbb{C}^{N_{\mathrm r} \times N_{\mathrm r}}$ and $\boldsymbol{\Lambda}_{{\rm T}}
\in\mathbb{C}^{N_{\mathrm t} \times N_{\mathrm t}}$ are diagonal matrices.

For this channel model, the optimal left singular matrix $\mathbf{U}_{{\mathrm{\bf B}}}$ of  precoder $\qB$
equals $\mathbf{U}_{{\rm T}}$ \cite{zeng2012linear}.
From this, using \cite[Eq. (5)]{zeng2012linear}, and
recalling (\ref{eq:x}) and (\ref{eq:Spatial_Cov}),  we can rewrite (\ref{eq:x}) as
\begin{equation}\label{eq:y_eq}
{{\bf{y}}_{{\rm{eq}}}} = {{\bf{H}}_{{\rm{eq}}}} \qx_{\rm eq}  + {\bf{\tilde{n}}}
\end{equation}
where
\begin{align}\label{eq:x_eq}
{{\bf{y}}_{{\rm{eq}}}} & =  \qU_{\rm R}^H \qy \\
\qx_{\rm eq} & =  { {{\bf{\Lambda }}_{{{\mathrm{\bf{B}}}}}}{{\bf{V}}_{{{\mathrm{\bf{B}}}}}} } \, \qd  \label{eq:x_eq}\\
{{\bf{H}}_{{\rm{eq}}}}& = \boldsymbol{\Lambda}_{{\rm R}}^{1/2} \tilde{\qW}  \boldsymbol{\Lambda}_{{\rm T}}^{1/2}
\end{align}
and where ${\bf{\tilde{n}}}$ and $\tilde{\qW}$ have the same distributions as $\qn$ in (\ref{eq:x}) and $\qW$ in (\ref{eq:Spatial_Cov}), respectively.

\subsection{Mutual Information in the Large-Dimensional Regime}
In order to obtain counterparts of (\ref{eq:y_4_exp_5}) and (\ref{eq:y_4_exp_6}) for this setting, we move into
 the large-dimensional regime \cite{Wu2015TWCOM}.
When both $N_{\mathrm r}$ and $N_{\mathrm t}$ grow
large with ratio $c = N_{\mathrm t}/N_{\mathrm r}$, the mutual information in (\ref{eq:Mutual_Info})
satisfies \cite{Wu2015TWCOM}
\begin{equation}\label{eq:GAUMutuall}
I(\qx;\qy) \simeq I_{\rm asy}(\qx;\qy)
\end{equation}
where
\begin{align}
 I_{\rm asy}(\qx;\qy) & = I\left( \qx_{\rm eq}  ;\qz_{\rm eq} \right) + \log_2 \det\left(\qI_{N_{\mathrm r}} + {\mathbf R}_{\rm eq} \right) \nonumber \\
& \quad -  \gamma_{\rm eq}  \psi_{\rm eq} \log_2 e.
\label{eq:GAUMutuall_2}
\end{align}
given the diagonal MIMO relationship
\begin{equation}\label{eq:EqScalGAUEach}
\qz_{\rm eq} =  \qXi_{\rm eq}^{1/2} \qx_{\rm eq} + \check{\bf n}
\end{equation}
where $\qx_{\rm eq}$ is given in (\ref{eq:x_eq}) and
$\check{\bf n} \in \bbC^{N_{\mathrm t}} $ is a standard complex Gaussian random vector.
The diagonal matrix $\qXi_{\rm eq}$ is a function of
auxiliary variables $\{ \gamma_{\rm eq}, \psi_{\rm eq}, {\mathbf R}_{\rm eq} \}$,
which are the solutions of the following set of coupled equations
\begin{align} \label{eq:qXi_2}
    \qXi_{\rm eq} &  = \gamma_{\rm eq} \boldsymbol{\Lambda}_{{\rm T}} \\
         {\mathbf R}_{\rm eq} & = \psi_{\rm eq} \boldsymbol{\Lambda}_{{\rm R}} \label{eq:qXi_3}   \\
\gamma_{\rm eq} & =\tr \left( \left(\qI_{N_{\mathrm r}}+\mathbf{R}_{{\rm eq}} \right)^{-1} \boldsymbol{\Lambda}_{{\rm R}} \right) \label{eq:Varsigma_k-MSE_MuKron3}\\
\psi_{\rm eq} & = \tr  \left( {\qOmega_{\rm eq}} \boldsymbol{\Lambda}_{{\rm T}} \right). \label{eq:Varsigma_k-MSE_MuKron4}
    \end{align}
Computing $\qXi_{\rm eq}$ requires finding $\{ \gamma_{\rm eq}, \psi_{\rm eq}, {\mathbf R}_{\rm eq} \}$ through fixed point
equations (\ref{eq:qXi_2})--(\ref{eq:Varsigma_k-MSE_MuKron4}).
The diagonal relationship in (\ref{eq:EqScalGAUEach}) does not relate to any physical channel, but is merely an instrument to obtain an asymptotic expression for the mutual information.
Nevertheless, we shall take advantage of this relationship.

Also necessary for later derivations is the minimum mean square error (MMSE)
estimate of $\qx_{\rm eq}$ based on (\ref{eq:EqScalGAUEach}), which is given by
\begin{equation}\label{eq:hatx_k}
 \hat\qx_{\rm eq} = E\left[\qx_{\rm eq}|\qz_{\rm eq} \right] .
\end{equation}
It will be convenient to define the
following MMSE matrix as the covariance of the error vector between the transmitted signal and its estimate,
\begin{equation} \label{eq:Omega_eq}
  \qOmega_{\rm eq} = E\left[ (\qx_{\rm eq} - \hat\qx_{\rm eq}) (\qx_{\rm eq} - \hat\qx_{\rm eq})^H \right] .
\end{equation}

\subsection{Precoder Structure}

Let us divide the transmit signal $\qd$ into $S$ streams.
Let the set $\left\{\ell_{1},\ldots,\ell_{N_{\mathrm t}}\right\}$ denote
a permutation of $\left\{1,\cdots,N_{\mathrm t}\right\}$ and
let ${\bf{\Lambda }}_s \in \mathbb{C}^{N_{\mathrm s} \times N_{\mathrm s}}$
and $\qV_s \in \mathbb{C}^{N_{\mathrm s} \times N_{\mathrm s}}$ denote a diagonal matrix
and a unitary matrix, respectively, for $s = 1,\ldots,S$.
${\bf{\Lambda }}_s $ and $\qV_s$ will be optimized later.
The goal of arranging these $S$ streams as in (\ref{eq:y_4_exp_5})
and (\ref{eq:y_4_exp_6}) prompts the following design steps:

\subsubsection{Structure of ${\bf{\Lambda}}_{\mathrm{\bf{B}}}$}

We define
\begin{align}\label{eq:Lambda_pair}
{\left[ {{{\bf{\Lambda }}_{{{\mathrm{\bf{B}}}}}}} \right]_{\ell_{j}\ell_{j}}} = \left[ {{{\bf{\Lambda}}_{s}}} \right]_{ii}
\end{align}
where $i = 1,\ldots,N_{\mathrm s}$, $s=1,\ldots,S$, and $j = (s- 1)N_{\mathrm s} + i$.
Under this structure, the $s$th stream is transmitted along the $\ell_{(s - 1)N_{\mathrm s} + 1},\ldots, \ell_{(s - 1)N_{\mathrm s} + N_{\mathrm s}}$
diagonal entries of $\qXi_{\rm eq}$.

\subsubsection{Structure of ${{\bf{V}}_{{{\mathrm{\bf{B}}}}}}$}

We define
\begin{align}\label{eq:V_pair}
& \!\!\!\!  \left[ {{{\bf{V}}_{{{\mathrm{\bf{B}}}}}}} \right]_{\ell_i \ell_j} =  \\
& \left\{ \begin{array}{l}
   {\left[ {{{\bf{V}}_{s}}} \right]_{mn}} \quad {\rm if} \ i = (s - 1) N_{\mathrm s} + m , \ j = (s - 1) N_{\mathrm s} + n   \\
0 \qquad \quad\;\; {\rm otherwise}
 \end{array} \right. \nonumber
\end{align}
where $m = 1,\ldots, N_{\mathrm s}$, $n = 1,\ldots,N_{\mathrm s}$, $s = 1,\ldots, S$,
$i =  1,\ldots, N_{\mathrm t}$, and $j = 1,\ldots, N_{\mathrm t}$. Under this structure,
for the $s$th stream the entries of $\qV_{s}$ map only to rows $\ell_{( s- 1) N_{\mathrm s} + 1},\ldots, \ell_{( s- 1)N_{\mathrm s} + N_{\mathrm s}}$ and columns $\ell_{( s - 1)N_{\mathrm s} + 1},\ldots, \ell_{(s - 1) N_{\mathrm s} + N_{\mathrm s}}$ of  ${{{\bf{V}}_{{{\mathrm{\bf{B}}}}}}} $.
This yields $S$ decoupled groups of streams at the receiver.

The design in (\ref{eq:y_4_exp_4}) is a specific instance of
(\ref{eq:V_pair}) with $\left\{\ell_{1},\cdots,\ell_{N_{\mathrm t}}\right\}
 = \left\{1,2,3,4\right\}$ and $S =2$. Recall how $(d_1,d_2)$ and
 $(d_3,d_4)$ are indeed decoupled in (\ref{eq:y_4_exp_5}) and (\ref{eq:y_4_exp_6}).

\subsubsection{Structure of $\qd_{s}$}

Finally, we let
\begin{align}\label{eq:d_pair}
\left[\qd_{s}\right]_{i} =  \left[\qd \right]_{\ell_{j}} .
\end{align}

It is noted that for the precoder design with
perfect instantaneous CSI, similar decoupled structures as in (\ref{eq:Lambda_pair})--(\ref{eq:d_pair})
are presented in \cite{Ketseoglou2015TWC} based on a per-group precoding technique.

\subsection{Precoder Optimization}

Based on (\ref{eq:Lambda_pair})--(\ref{eq:d_pair}), the relationship in (\ref{eq:x_eq}) can be rewritten as
\begin{align}\label{eq:x_eq_pair}
 \left[\qx_{\rm eq}\right]_{\ell_{j}} =  \left[{\bf{\Lambda }}_{{s}}  {\bf{V}}_{{s}}  \qd_{{s}}\right]_{i}
\end{align}
for $i = 1,\ldots,N_{\mathrm s}$, $s = 1,\ldots, S$,  and  $j = ( s- 1)N_{\mathrm s} + i$.
Recalling that
$\qXi_{\rm eq}$ is diagonal, (\ref{eq:EqScalGAUEach}) then reduces to
\begin{align}\label{eq:z_eq_pair}
 \left[\qz_{\rm eq}\right]_{\ell_{j}} & = \left[ \qXi_{\rm eq} \right]_{\ell_{j}\ell_{j}}  \left[\qx_{\rm eq}\right]_{\ell_{j}} + [\check{\bf n}]_{\ell_{j}} .
\end{align}

Eqs. (\ref{eq:x_eq_pair}) and (\ref{eq:z_eq_pair}) indicate that each independent data stream
$\qd_{s} $ is transmitted along its own $N_{\mathrm s}$ separate subchannels without
interfering with other streams.
Furthermore,  the MMSE matrix in (\ref{eq:Omega_eq}) then equals
\begin{align}\label{eq:Omega_pair}
& \!\! {\left[  \qOmega_{\rm eq} \right]_{\ell_i \ell_j}} = \\
&  \left\{ \begin{array}{lll}
{\left[ \qOmega_{s}  \right]_{mn}} \;\; \ {\rm if} \ i = ( s - 1)N_{\mathrm s} + m , \ j = ( s- 1)N_{\mathrm s} + n   \\
0 \qquad \quad \;\, {\rm otherwise}   \\
\end{array} \right.
\end{align}
where
\begin{equation}\label{eq:MMSE_eq}
\qOmega_{s}  = {{\bf{\Lambda }}_{s}}{{\bf{V}}_{s}} \, E\left[ {\big( { {{\bf{d}}_{s}} - {{{\bf{\hat d}}}_{s}}} \big){{\big( {{{\bf{d}}_{s}} - {{{\bf{\hat d}}}_{s}} } \big)}^H}} \right] {\bf{V}}_{s}^H{\bf{\Lambda }}_{s}^H
\end{equation}

\begin{alg} \label{Gradient_Pair}
Maximization of $I\left( \qx;\qy \right) $ with respect to $\qB$.

\vspace*{1.5mm} \hrule \vspace*{1mm}
  \begin{enumerate}

\itemsep=0pt

\item Initialize $\mathbf{\Lambda}_{{s}}^{(0)}$, $\mathbf{V}_{{s}}^{(0)}$ for  $s = 1,\ldots,S$.
Fix a maximum number of iterations, $N_{\rm iter}$, and a threshold $\varepsilon$.

\item  Initialize $\qXi_{\rm eq}$, ${\bf{R}}_{\rm eq}$,
 $\gamma_{\rm eq}$, and $\psi_{\rm eq}$
based on (\ref{eq:qXi_2})--(\ref{eq:Varsigma_k-MSE_MuKron4}), with $\qOmega_{\rm eq}$
based on (\ref{eq:Omega_pair}).  Then, initialize $I^{\left(1\right)}\left( \qx;\qy \right)$ based on (\ref{eq:GAUMutuall}) with
 $I\left( \qx_{\rm eq} ;\qz_{\rm eq} \right)$ as per (\ref{eq:I_pair}).  Set counter $n = 1$.

\item Update ${\mathbf{\Lambda}}_{{s}}^{(n)}$ for $s= 1,\ldots,S$ along
the gradient descent direction given by (\ref{eq:gamma_gradient}).

\item Normalize $ \sum\nolimits_{s=1}^{S} \tr \left(\left({\mathbf{\Lambda}}_{{s}}^{(n)} \right)^2 \right)= P$.

\item Update $\mathbf{V}_{{s}}^{(n)}$ for $s= 1,\ldots,S$ along the gradient descent direction in (\ref{eq:V_gradient}).

\item Update $\qXi_{\rm eq}$, ${\bf{R}}_{\rm eq}$,
 $\gamma_{\rm eq}$, and $\psi_{\rm eq}$
based on (\ref{eq:qXi_2})--(\ref{eq:Varsigma_k-MSE_MuKron4}), (\ref{eq:Omega_pair}).

\item Compute $I^{\left(n + 1\right)}\left( \qx;\qy \right)$ based on (\ref{eq:GAUMutuall}) and (\ref{eq:I_pair}).
If $I^{\left(n + 1\right)}\left( \qx;\qy \right)  - I^{\left(n  \right)}\left( \qx;\qy \right) > \varepsilon$
and $n \leq N_{\rm iter}$,  set $n = n + 1$ and repeat Steps $3$--$7$;

\item Compute ${{{\bf{\Lambda }}_{{{\mathrm{\bf{B}}}}}}}$ and
 ${{{\bf{V}}_{{{\mathrm{\bf{B}}}}}}}$ based on (\ref{eq:Lambda_pair}) and (\ref{eq:V_pair}).  Set $\qB = \mathbf{U}_{{\rm T}} {{{\bf{\Lambda }}_{{{\mathrm{\bf{B}}}}}}}
{{{\bf{V}}_{{{\mathrm{\bf{B}}}}}}}$.

 \vspace*{1mm} \hrule

  \end{enumerate}

\end{alg}
\null
\par

\noindent with

\begin{equation}\label{eq:q}
{{{\bf{\hat d}}}_{s}} = E\left[ {{\bf{d}}_{s}} \Big|\qz_{s} \right]
\end{equation}
and $\left[\qz_{s}\right]_{i} =  \left[\qz_{\rm eq}\right]_{\ell_{j}}$.


The main term in the mutual information in (\ref{eq:GAUMutuall}) is $I\left( \qx_{\rm eq} ;\qz_{\rm eq} \right)$, which can now be expressed as
\begin{equation}\label{eq:I_pair}
I\left( \qx_{\rm eq} ;\qz_{\rm eq} \right) = \sum\limits_{s = 1}^S I\left( {{\bf{d}}_{s}};\qz_{s}  \right)
\end{equation}
based on which the gradients of $ I_{\rm asy}(\qx;\qy)$ with respect to
$\mathbf{\Lambda}_{s}^2$ and $\mathbf{V}_{s}$ are given by \cite[Eq. (22)]{Palomar2006TIT},
\begin{align}
\label{eq:gamma_gradient}
{\nabla _{{\bf{\Lambda }}_{s}^2}} I_{\rm asy}(\qx;\qy) & =  {\rm diag} \left( {{\bf{V}}_{s}^H{{\bf{E}}_{s}}{{\bf{V}}_{s}}} \qXi_{s} \right) \\
\label{eq:V_gradient}
{\nabla _{{{\bf{V}}_{s}}}} I_{\rm asy}(\qx;\qy) &  =   \qXi_{s} {\bf{\Lambda }}_{s}^2{{\bf{V}}_{s}}{{\bf{E}}_{s}}
\end{align}
where
\begin{equation}\label{eq:Es}
{{\bf{E}}_{s}} = E\left[ {\left( { {{\bf{d}}_{s}} - {{{\bf{\hat d}}}_{s}}} \right){{\left( {{{\bf{d}}_{s}}
- {{{\bf{\hat d}}}_{s}} } \right)}^H}} \right]
\end{equation}
and we define diagonal matrices $\qXi_s$, for $s=1,\ldots,S$, with entries $\left[\qXi_{s}\right]_{ii} = \left[ \qXi_{\rm eq} \right]_{\ell_{j}\ell_{j}}$.

From (\ref{eq:I_pair}), and from the relationship between $\mathbf{\Lambda}_{1},\ldots,\mathbf{\Lambda}_{S}$ and ${{{\bf{\Lambda }}_{{{\mathrm{\bf{B}}}}}}}$ in (\ref{eq:Lambda_pair}) as well as the relationship between
$\mathbf{V}_{1},\ldots,\mathbf{V}_{S}$ and ${{{\bf{V}}_{{{\mathrm{\bf{B}}}}}}}$ in (\ref{eq:V_pair}),
we propose Algorithm \ref{Gradient_Pair} to optimize ${{{\bf{\Lambda }}_{{{\mathrm{\bf{B}}}}}}}$ and ${{{\bf{V}}_{{{\mathrm{\bf{B}}}}}}}$.
In Steps 3 and 5 of Algorithm 1, ${\mathbf{\Lambda}}_{{s}}^{(n)}$ and $\mathbf{V}_{{s}}^{(n)}$
are updated along the gradient descent direction, with the backtracking line search method used to determine the step size.
In Step 4, ${\mathbf{\Lambda}}_{{s}}^{(n)}$ is normalized to satisfy the power constraint.
In Step 6, $\qXi_{\rm eq}$, ${\bf{R}}_{\rm eq}$,
 $\gamma_{\rm eq}$, and $\psi_{\rm eq}$ are updated for the new precoder based on
(\ref{eq:qXi_2})--(\ref{eq:Varsigma_k-MSE_MuKron4}), (\ref{eq:Omega_pair}).
In Step 7, if  $n$ is less than the maximal number of iterations
and $I^{\left(n + 1\right)}\left( \qx;\qy \right)  - I^{\left(n  \right)}\left( \qx;\qy \right)$
is larger than a threshold, we implement the next iteration, otherwise, we compute the final
precoder and stop the algorithm.

\textit{Remark 1:} For the complete-search design algorithm \cite{Xiao2011TSP,zeng2012linear}, the complexity is dominated by the
computation of the mutual information and the MMSE matrix, which grows exponentially with $2 N_{\mathrm t}$.
For Algorithm \ref{Gradient_Pair}, alternatively, the complexity of computing the mutual information and the MMSE matrix in Algorithm \ref{Gradient_Pair} grows exponentially with $2 N_{\mathrm s}$.
Thus, by choosing proper values of $S$ and $N_{\mathrm s}$, Algorithm \ref{Gradient_Pair} offers a tradeoff between performance and complexity.
At one end, when $S =1$ and $N_{\mathrm s} = N_{\mathrm t}$, Algorithm \ref{Gradient_Pair}
searches the entire space, while at the other end,
when $S = N_{\mathrm t}$ and $N_{\mathrm s} = 1$, Algorithm \ref{Gradient_Pair} merely allocates power among the $N_{\mathrm t}$ parallel subchannels.
Varying $N_{\mathrm s}$ from $1$ to $N_{\mathrm t}$
bridges the gap between separate and fully joint transmission of the $N_{\mathrm t}$ original signals.

\textit{Remark 2:}
An adequate choice of $\ell_{1},\ldots, \ell_{N_{\mathrm t}}$ is important for Algorithm \ref{Gradient_Pair}
to perform satisfactorily. As discussed in Section III, the most important step to
compensate the performance loss caused by the Gaussian input design is to pair a strong
subchannel with a weak subchannel and transmit the mixed signals along them together.
Therefore, the $N_{\mathrm s}/2$ largest diagonal entries of $[\qXi_{\rm eq}]$ are paired with the $N_{\mathrm s}/2$ smallest diagonal entries.
Then, the remaining $N_{\mathrm s}/2$ largest diagonal entries of $[\qXi_{\rm eq}]$ are paired with
the remaining $N_{\mathrm s}/2$ smallest ones, and so on.
This generalizes the two-antenna scheme in \cite{Mohammed2011TIT}.

\textit{Remark 3:} Since ${\mathbf{\Lambda}}_{{s}}^{(n)}$ and $\mathbf{V}_{{s}}^{(n)}$
are searched along the gradient descent direction, in Step 7 the mutual information $I^{\left(n \right)}\left( \qx;\qy \right)$ is nondecreasing.
Since Algorithm \ref{Gradient_Pair} generates sequences that are nondecreasing and upper-bounded, it is convergent.
However, due to the nonconvexity of $I^{\left(n \right)}\left( \qx;\qy \right)$ in ${\mathbf{\Lambda}}_{s}^{(n)}$ and $\mathbf{V}_{s}^{(n)}$, Algorithm \ref{Gradient_Pair} may only find local optima.
As a result, the algorithm is run several times with different
random initializations of ${\mathbf{\Lambda}}_{{s}}^{(n)}$ and $\mathbf{V}_{{s}}^{(n)}$ and the final precoder that provides the highest
mutual information is retained.

\section{Performance Evaluation}


First, let us evaluate the complexity of Algorithm 1 for different values of $N_{\mathrm s}$.
Matlab is used on an Intel Core i7-4510U 2.6 GHz processor.
Tables I--III provide the run time per iteration, for various numbers of antennas and constellations,
with $\times$ indicating that the time exceeds one hour. As predicted, for $N_{\mathrm s} = N_{\mathrm t}$, the computational complexity grows exponentially with $N_{\mathrm t}$ and quickly becomes unwieldy.

\begin{table}[!t]

\label{runing_time_1}
\centering
 \captionstyle{center}
  {
\caption{Run time (sec.) per iteration with BPSK. }
\begin{tabular}{|c|c|c|c|}
\hline
  $N_{\mathrm t}$  &     $ N_{\mathrm s} = 2$   &   $ N_{\mathrm s} = 4$  &  $N_{\mathrm s} = N_{\mathrm t}$        \\ \hline
  $4$  & 0.0051   &   0.0190   &  0.0190   \\ \hline
 $8$    &  0.0112   &  0.0473  &  11.6209       \\ \hline
   $16$    &  0.0210    &   0.1939 &   $\times$     \\ \hline
      $32$    &  0.0570    &   0.4111 &   $\times$      \\ \hline
\end{tabular}
}
\end{table}

\begin{table}[!t]

\label{runing_time_2}
\centering
\captionstyle{center}
  {
\caption{Run time (sec.) per iteration with QPSK.}
\begin{tabular}{|c|c|c|c|}
\hline
  $N_{\mathrm t}$  &     $ N_{\mathrm s} = 2$   &   $ N_{\mathrm s} = 4$  &  $N_{\mathrm s} = N_{\mathrm t}$        \\ \hline
  $4$  &  0.1149 &   21.5350   &  21.5350  \\ \hline
 $8$    &  0.2029   &   23.3442  &  $\times$        \\ \hline
   $16$    & 0.3001   &   48.1725 &   $\times$     \\ \hline
      $32$    & 0.7094   &   98.7853&   $\times$      \\ \hline
\end{tabular}
}
\end{table}

\begin{table}[!t]

\label{runing_time_3}
\centering
 \captionstyle{center}
  {
\caption{Run time (sec.) per iteration with 16-QAM.}
\begin{tabular}{|c|c|c|}
\hline
  $N_{\mathrm t}$  &     $ N_{\mathrm s} = 2$    &  $N_{\mathrm s} = N_{\mathrm t}$        \\ \hline
  $4$  &  28.0744    &   $\times$  \\ \hline
 $8$    &  58.3433 &     $\times$        \\ \hline
   $16$    & 106.6022      &   $\times$     \\ \hline
      $32$    & 233.2293 &   $\times$      \\ \hline
\end{tabular}
}
\end{table}

\begin{figure}[!t]
\centering
\includegraphics[width=0.5\textwidth]{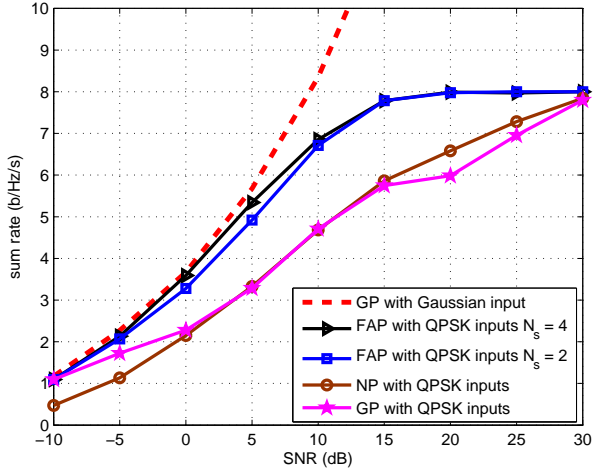}
 \captionstyle{flushleft}
 \caption{Spectral efficiency versus SNR for the 3GPP SCM for different precoder designs with $N_{\mathrm t} = N_{\mathrm r} =4$ and QPSK.}
\label{QPSK_SCM}
\end{figure}

Figure \ref{QPSK_SCM} depicts the spectral efficiency for the 3GPP SCM (urban scenario, half-wavelengh antenna spacing, velocity $36$ km/h) for different precoder designs with $N_{\mathrm t} = N_{\mathrm r} = 4$ and QPSK.
A Gauss-Seidel algorithm using stochastic programming is employed
to obtain the capacity-achieving precoder \cite{wen2011sum}. For
Algorithm \ref{Gradient_Pair}, both $N_{\mathrm s} = 4$ and $N_{\mathrm s} = 2$ are considered, and despite their enormous computational gap (cf. Table II) the difference in performance is minor.
Both precoders hug the capacity up to the point where the QPSK cardinality becomes insufficient.
The precoder designed via Algorithm \ref{Gradient_Pair}
gains many dBs over an unprecoded transmitter and also over a capacity-achieving precoder applied with QPSK.

\begin{figure}[!t]
\centering
\includegraphics[width=0.5\textwidth]{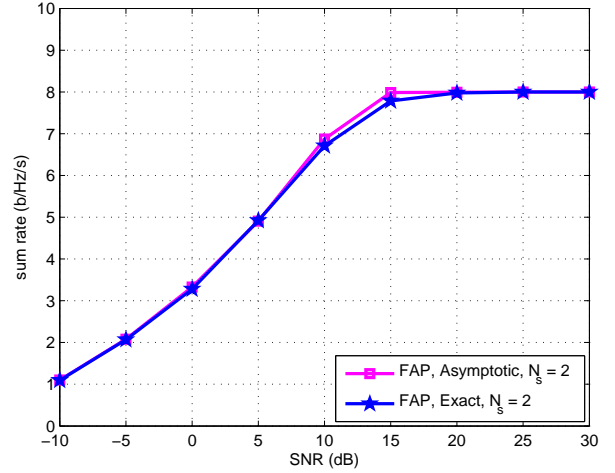}
 \captionstyle{flushleft}
\caption{Asymptotic and exact spectral efficiency versus SNR for the 3GPP SCM (urban scenario, half-wavelength antenna spacing, $36$ km/h) with $N_{\mathrm t} = N_{\mathrm r} =4$ and QPSK.}
\label{QPSK_SCM_asy}
\end{figure}

Figure \ref{QPSK_SCM_asy} contrasts the spectral efficiency given by the asymptotic expression in (\ref{eq:GAUMutuall}) with the exact form in (\ref{eq:Mutual_Info}) for the precoders obtained
by Algorithm \ref{Gradient_Pair} with $N_{\mathrm s} = 2$. The channel model is the same as for Fig. \ref{QPSK_SCM}.
The perfect match between the two curves confirms that (\ref{eq:GAUMutuall}) is a very good proxy for (\ref{eq:Mutual_Info}), and hence that Algorithm 1 is indeed effective even for small numbers of antennas.

\begin{figure}[!t]
\centering
\includegraphics[width=0.5\textwidth]{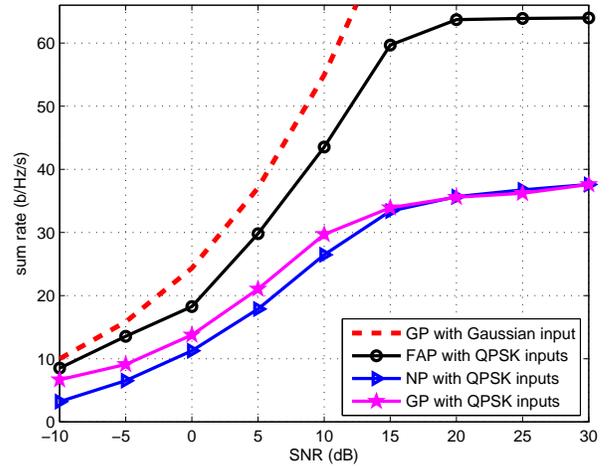}
 \captionstyle{flushleft}
\caption{Spectral efficiency versus SNR for the 3GPP SCM for different precoder designs with $N_{\mathrm t} = N_{\mathrm r} =32$, $N_s = 4$, and QPSK.}
\label{QPSK_SCM_32}
\end{figure}

\begin{figure}[!t]
\centering
\includegraphics[width=0.5\textwidth]{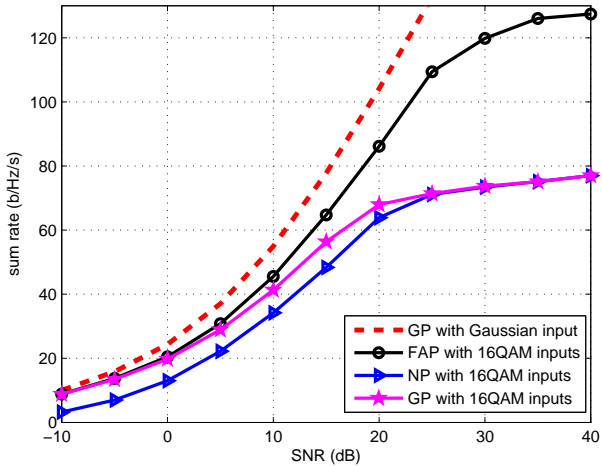}
 \captionstyle{flushleft}
\caption{Spectral efficiency versus SNR for the 3GPP SCM for different precoder designs with $N_{\mathrm t} = N_{\mathrm r} =32$, $N_s = 2$, and 16-QAM.}
\label{16QAM_SCM_32}
\end{figure}

Figures \ref{QPSK_SCM_32} and \ref{16QAM_SCM_32} present further results for the same SCM parameter settings
as in Figure \ref{QPSK_SCM} and \ref{QPSK_SCM_asy} for $N_{\mathrm t} = N_{\mathrm r} = 32$ and for QPSK and 16-quadrature amplitude modulation (QAM), respectively.  We set $N_{\mathrm s} =4$ for the former and $N_{\mathrm s} = 2$ for the latter.
We note that precoder design for such large arrays were, to best of the authors' knowledge, not available henceforth for discrete signals (except for \cite{Ketseoglou2016}, which was available online after the submission of this paper).
As the numbers of antennas grow, the conditioning of the transmit correlation matrix $\qA_{{\rm T}}$ becomes progressively poorer \cite{Adhikary2013TIT} and the performance of the capacity-achieving precoder applied to discrete signals degrades, even failing to achieve the saturation spectral efficiency of $N_{\mathrm t} \log_2 M$ b/s/Hz at relevant SNRs; some subchannels are simply never activated by a precoder intended for Gaussian signals.
Algorithm 1, in contrast, is tailored to finite-cardinality constellations.

\section{Conclusion}

With a proper design of $\mathbf{V}_{{\mathrm{\bf{B}}}}$ (right unitary matrix in the SVD decomposition of the precoder), it is possible to achieve a satisfactory tradeoff between the need to feed into the channel mixings of multiple finite-cardinality signals and the computational complexity of exploring all possible such mixings. Building on this idea, an algorithm has been proposed that---under the 3GPP SCM channel model---exhibits very good performance with orders-of-magnitude less complexity than complete-search solutions. More refined versions of this algorithm, equipped with alternative subchannel pairing schemes, may perform even better. Additional extensions include the applicability to settings with imperfect CSI, or to multiuser contexts, as well as the performance under other channel models.



\end{document}